\documentclass{article}

\PassOptionsToPackage{authoryear,round}{natbib}

\usepackage[preprint, position]{neurips_2026}

\usepackage[utf8]{inputenc}
\usepackage[T1]{fontenc}
\usepackage{url}
\usepackage{booktabs}
\usepackage{amsmath}
\usepackage{amsfonts}
\usepackage{nicefrac}
\usepackage{microtype}
\usepackage{xcolor}
\usepackage[colorlinks=true,linkcolor=black,citecolor=green!50!black,urlcolor=blue]{hyperref}
\usepackage{graphicx}
\usepackage{tabularx}
\usepackage{array}
\usepackage{enumitem}

\title{Metacognition Should Be the Scientific Framework for Bounded and Effective Self-Governance in Generative AI}

\author{
Eugene Yu Ji \\
University of Waterloo \\
\texttt{eugeneyuji1@gmail.com}
\And
Igor Grossmann \\
University of Waterloo \\
\texttt{igrossma@uwaterloo.ca}
\And
Amir-Hossein Karimi \\
University of Waterloo \\
\texttt{amirh.karimi@uwaterloo.ca}
}

\begin{document}
\maketitle

\begin{abstract}
Generative AI research increasingly confronts a shared problem: systems must sustain yet govern their own generative activity when uncertainty is high, evidence is missing, or context is insufficient. This position paper argues that \textbf{metacognition should become the scientific framework for bounded and effective self-governance in generative AI}, where output generation is properly evaluated together with the capacities through which generative systems navigate and regulate their own activity. We advance this position by showing that bounded and effective AI self-governance requires metacognitive alignment across computational, algorithmic, and ecological levels. At the \textit{computational} level, metacognition specifies the meta-level functions a system is meant to serve, such as monitoring, evaluation, control, and adaptation. At the \textit{algorithmic} level, these functions are realized through procedures such as elicitation, iteration, and modularization. At the \textit{ecological} level, metacognitive signals become meaningful, actionable, and accountable within the interface, workflow, and accountability arrangements. Metacognition thus makes it possible to conceive generative AI as both capable and well-governed, rather than treating capability and governance as competing aims.\end{abstract}

\section{Introduction}

Generative AI increasingly requires systems whose task- and object-level activities can be regulated by ``meta-level'' strategies \citep{johnson2026}.  We argue that \textbf{metacognition should be the scientific framework for bounded and effective self-governance in generative AI}, where generation is explained and evaluated together with the capacities through which generative systems manage their own activity. 

The term ``metacognition'' comes from psychology, where it refers broadly to knowledge, awareness, monitoring, and control directed toward one's own mental activity \citep{flavell1979,nelson1990}. Computer science provides a related lineage: early ``metareasoning'' studies the cost, value, and control of computation under bounded resources \citep{russell1989,russell1991}, while related contemporary work extends this concern to introspective monitoring, meta-level control, anomaly detection, and failure recovery \citep{cox2005,schmill2011,cox2011}. Although metacognition offers an important shared vocabulary for developing generative AI systems that can assess, update, regulate, and govern the adequacy and effectiveness of their own behavior, the term becomes conceptually and analytically unstable when applied across heterogeneous lines of work. In contemporary AI research, ``metacognitive'' can refer to psychology-inspired prompting \citep{wang2024}, retrieval planning \citep{zhou2024}, learning \citep{liu2025}, confidence calibration and verbalized uncertainty \citep{lin2022,tian2023,xiong2024}, internal-state monitoring \citep{azaria2023,jian2025}, tool-use control \citep{schick2023,yao2023b}, verifier-guided correction \citep{dhuliawala2024,gou2024}, agentic memory and reflection \citep{shinn2023,zhao2024}, or broader aspirations for safer, more governable AI systems \citep{bai2022,johnson2026}. 

We contend that these are fragments of a single scientific problem: how generative systems can achieve proper self-governance over their own generative activity. This paper argues that the goal should be \textbf{bounded \textit{and} effective self-governance}, and that this goal is impossible to specify, design, or evaluate without \textbf{cross-level metacognitive alignment}, in which \textit{functions}, \textit{procedures}, and \textit{ecological} conditions are precisely distinguished and systematically related, extending Marr's well-known three-level analysis of intelligent systems \citep{marr1982,krafft2018}.

\section{What Metacognition Can Mean for Self-Governing AI}

Metacognition entered cognitive psychology through models of monitoring and control of task-level processes that perform the primary activity by meta-level processes that observe, evaluate, and guide that activity \citep{flavell1979,nelson1990}. Later research shows that such monitoring is shaped by development, affect, embodiment, social interaction, and context, and can dissociate from task performance \citep{metcalfe1994,veenman2006,efklides2008,efklides2011,dunlosky2009,fleming2012,yeung2012,grossmann2017,grossmann2020,grossmann2021}. This lineage is helpful for conceptualizing AI self-governance, but must be translated carefully: metacognition for self-governance should refer to the functional capacity to monitor, evaluate, control, and adapt one's own activity, not to conscious introspection, subjective awareness, or human-like access to mental states \citep{overgaard2012,proust2013}. 

A similar computer-science tradition, from rational metareasoning to computational metacognition, treats intelligent systems as monitoring and controlling their own reasoning \citep{russell1989,russell1991,cox2005,cox2011,schmill2011,cox2022,caro2022}. Recent work extends this idea to generative AI through metacognitive prompting, retrieval monitoring and planning, activation-level self-report, and calibration-based assessment \citep{wang2024,zhou2024,jian2025,liu2025,servajean2026}. Additionally, a broader literature in contemporary computer science also develops procedures related to self-governance without using the language of metacognition. These include meta-level selection over reasoning paths \citep{wang2023,yao2023a}, feedback-based revision and memory \citep{madaan2023,shinn2023,gao2023,gou2024}, and coordination with retrieval, tools, or external action \citep{lewis2020,schick2023,yao2023b,zhou2024}. Work on self-correction makes the same issue explicit: models may assess confidence, limits, and the need for assistance, but such correction remains fragile and often depends on prompts, verifiers, or external scaffolding \citep{lin2022,kadavath2022,ren2023,xiong2024,kapoor2024,huang2024,li2024}.

These developments converge on the target of our position: the insufficiency of a capability-first view of generative AI \citep{johnson2026}. This view prioritizes task-level capabilities, such as fluent generation, answer accuracy, problem-solving performance, planning quality, and successful task completion, while treating meta-level capacities, such as uncertainty monitoring, sufficiency evaluation, self-correction, perspective-conditioned reasoning, as auxiliary or post-hoc additions to model performance. Against this view, we argue that bounded and effective self-governance requires task-level capacities to be properly organized by metacognitive capacities.

\section{Cross-Level Metacognitive Alignment for Self-Governing AI}

Our proposal is that bounded and effective self-governance in generative AI requires cross-level metacognitive alignment: \textit{computational targets}, \textit{algorithmic procedures}, and \textit{ecological arrangements} must be specified together rather than treated as separable design problems. By \textbf{bounded} self-governance, we mean that generative activity is not only treated as capability-driven open-ended production, but proceeds under metacognitive conditions that specify what should be monitored and evaluated, how control and adaptation should be procedurally realized, and where their signals and actions become meaningful in use. By \textbf{effective} self-governance, we mean that these conditions do not merely constrain generation, but make task-level activity more reliable, useful, and accountable. 

\begin{quote}
    \textbf{Working definition:} Metacognition in a generative system refers to self-governing processes through which the system uses information about its own states, outputs, and trajectories to manage subsequent generative behavior. At the \textbf{computational} level, metacognition specifies what should be \textit{monitored} and \textit{evaluated} and what forms of \textit{control} or \textit{adaptation} these assessments are meant to guide. At the \textbf{algorithmic} level, these functions are realized through procedures such as\textit{ elicitation},\textit{ iteration}, and\textit{ modularization}. At the \textbf{ecological} level, they are situated within \textit{interface}, \textit{workflow}, and \textit{accountability} arrangements that determine whether metacognitive signals and actions become meaningful, actionable, and accountable in use.
\end{quote}

\begin{figure}[t]
\centering
\includegraphics[width=1.1\linewidth]{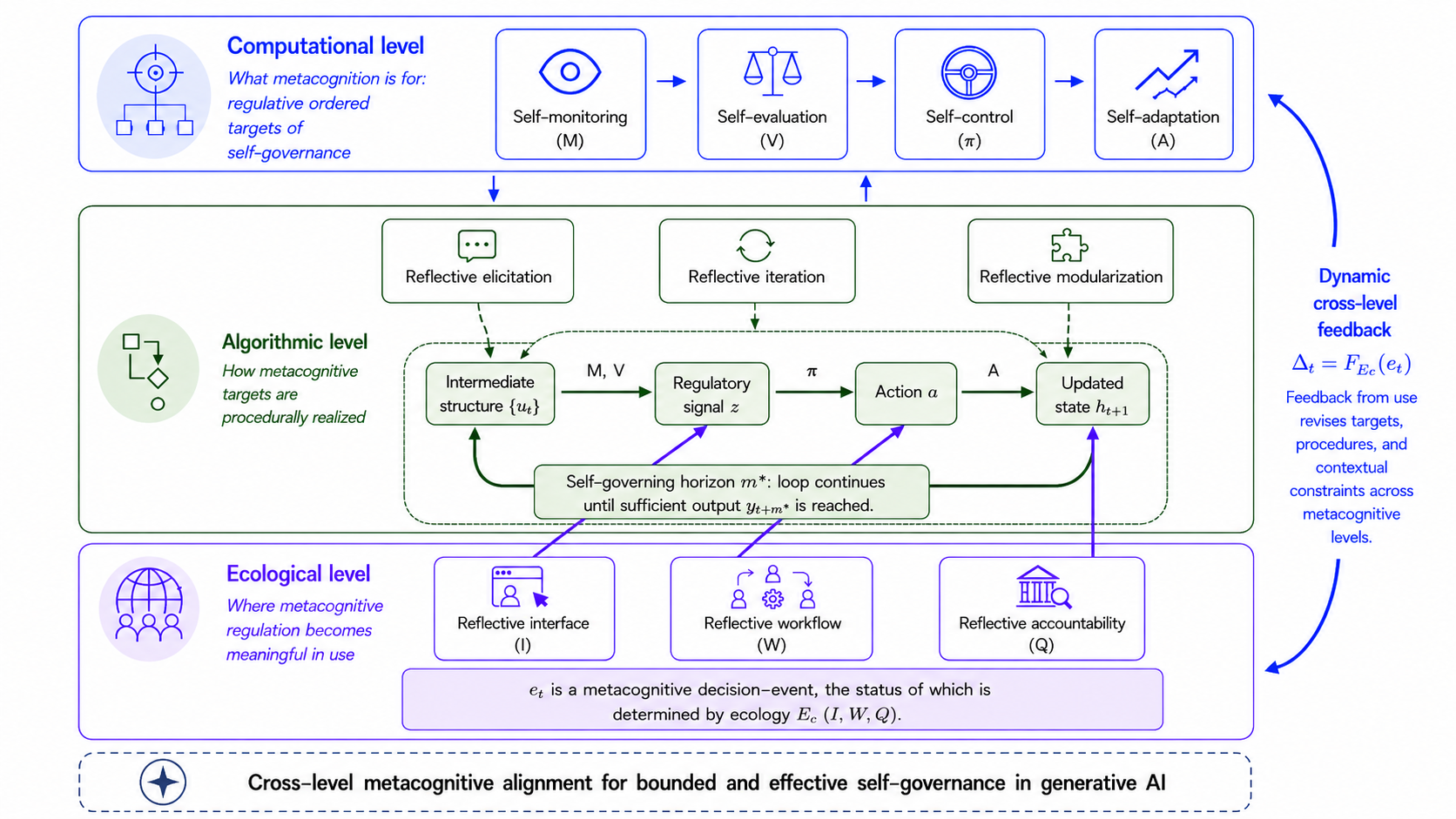}
\caption{Cross-level metacognitive alignment as a tripartite framework for bounded and effective self-governance in generative AI. The framework distinguishes computational targets, algorithmic realizations, and ecological conditions as three levels of metacognitive regulation, related through dynamic cross-level feedback. }
\label{fig:framework}
\end{figure}

This definition treats bounded and effective self-governance as cross-level metacognitive alignment rather than as a property of single-level or single-model capability (\hyperref[fig:framework]{Figure~\ref*{fig:framework}}). The tripartite-level approach takes loose inspiration from Marr's distinction among computational, algorithmic, and implementational levels for intelligent systems \citep{marr1982}, along with later work extending or adapting levels of analysis in cognitive science, machine learning, and LLMs \citep{griffiths2015,bechtel2015,krafft2018,hamrick2020,ku2025}. Our use of levels is deliberately revisionary. Marr's framework primarily differentiates explanatory levels and often treats the computational level as the privileged site for specifying the problem a system solves. Our framework treats differentiation and interrelations as equally central. In particular, we replace implementation with ecology to capture how metacognitive governance is realized under real sociotechnical conditions, both within and across models, prompts, tools,  users, workflows, organizations, and institutions \citep{hutchins1995,hollan2000,selbst2019,suresh2021}. 

\subsection{The computational level}

At the computational level, the central question is what metacognition is for: what meta-level problem the system is meant to solve with respect to its own generative activity. We treat \textit{self-monitoring}, \textit{self-evaluation}, \textit{self-control}, and \textit{self-adaptation} as regulatively ordered targets for self-governance: monitoring makes aspects of generation available; evaluation assigns significance to them; control changes the immediate course of generation; and adaptation carries regulatory consequences forward. Existing work already addresses certain important parts of these problems. Yet these contributions usually remain partial or loosely connected. 

\textbf{Self-monitoring}. Self-monitoring tracks conditions of the system's own activity that may matter for later regulation. In generative AI,  these conditions include analogous monitoring problems in verbalized uncertainty \citep{lin2022,xiong2024}, confidence calibration \citep{tian2023}, semantic uncertainty \citep{kuhn2023}, internal-state signals of truthfulness or epistemic condition \citep{azaria2023,jian2025}, among others. These studies make monitoring visible as a computational target, but their limit is that monitoring is often treated as an endpoint. Monitoring is metacognitive only when the tracked condition is available for subsequent evaluation or control, rather than being largely reported as an isolated diagnostic.

\textbf{Self-evaluation}. Self-evaluation assigns regulatory significance to what has been monitored. A monitored state becomes evaluative when uncertainty is treated as a reliability, risk, or self-improving signal \citep{lin2022,kadavath2022,tian2023,kuhn2023,xiong2024,kapoor2024,liu2025}; when inconsistency across reasoning paths guides selection or search \citep{wang2023,yao2023a}; when retrieval sufficiency is accessed \citep{lewis2020,gao2023,zhou2024}; or when verifier disagreement triggers correction or further validation \citep{dhuliawala2024,gou2024}. 

\textbf{Self-control}. Self-control turns evaluative signals into changes in the immediate trajectory of generation. The system may resample, expand, defer, abstain, or prune a reasoning path, retrieve evidence, call a tool, revise an answer. This target connects classical metareasoning on bounded computation and expected-value control \citep{russell1989,russell1991}, with contemporary generative AI methods for inference-time search \citep{yao2023a}, retrieval planning and repair \citep{zhou2024}, tool invocation \citep{schick2023,yao2023b}, and selective answering or abstention under uncertainty \citep{ren2023,kapoor2024}. 

\textbf{Self-adaptation}. Self-adaptation carries the consequences of regulation into later behavior. It concerns how systems maintain or adjust their capacities, limits, memories, strategies, and conditions of success over time \citep{cox2005,schmill2011,cox2011,cox2022}. In generative AI, this most often appears in reflective memory \citep{shinn2023,zhao2024}, experiential learning from task histories \citep{cox2022,zhao2024,liu2025}, failure recovery and feedback-guided strategy revision \citep{madaan2023,dhuliawala2024,gou2024}. 

Together, these four interconnected targets define metacognition at the computational level. They specify the reflective navigating and regulatory problem that bounded and effective generative activity must solve: how the system tracks its epistemic and operational condition, interprets the significance of that condition, manages the action according to the condition, and reshapes later behavior in light of them.

\subsection{Algorithmic level}

At the algorithmic level, we define a minimal algorithmic template of how metacognitive computational targets become proper self-governing procedures. Let $x$ denote the input task, $c$ the default task context, $h_t$ the relevant system state, $G$ the base generative model, and $y$ the generated output:
\[
y \sim G(\cdot \mid x,c,h_t).
\]
Metacognition-relevant algorithms introduce intermediate structure $u$, such as rationales, decompositions, sampled trajectories, uncertainty estimates, retrieved evidence, verifier scores, memory traces, perspective taking, and so on:
\[
u = G_{\mathrm{int}}(x,c,h_t).
\]
Here $u$ as $G_{\mathrm{int}}$ becomes metacognitively relevant only when it is interpreted as a regulatory signal $z$, mapped by a control policy $\pi$ into an action $a$, and, in stronger cases, carried forward through adaptation $A$:
\[
u \xrightarrow{M,V} z \xrightarrow{\pi} a \xrightarrow{A} h_{t+1}.
\]
A self-regulatory signal $z$ is an interpreted assessment of the state of the generation under monitoring $M$ and evaluation $V$. A control policy $\pi$ maps such signals to regulatory actions $a$. Adaptation $A$ carries the consequences of regulation into the later system state $h_{t+1}$. The algorithmic question is therefore whether $z$ is diagnostically valid and $\pi$ routes it to actions that improve reliability, usefulness, or accountability (effectiveness) and how the action $a$ regulates the later system state $h_{t+1}$ specifies (boundedness). The algorithmic question is to determine a proper \textbf{self-governing horizon }$m^\ast$ such that \[
m^\ast = \inf \{ k \geq 0 : y_{t+k} \text{ is sufficient for } g(x,c) \text{ under } L(c,h_{t+k}) \}.
\]
Here $g(x,c)$ is the task-level goal, and $L(c,h_{t+k})$ denotes the relevant metacognitive constraints in task context $c$. In other words, the system seeks an output $y_{t+m^\ast}$ that is \textit{effective} for achieving a task goal $g$ in context $c$, while regulating the conditions under which that output remains \textit{bounded}, with the latter conditions characterized through the actual ecological characteristics of context $c$.

\textbf{Reflective elicitation}. Reflective elicitation makes intermediate structure $u=G_{\mathrm{int}}(x,c,h_t)$ available as a possible object of monitoring and evaluation, through which the system can estimate whether it is close to, far from, or wrongly oriented toward $m^\ast$:
\[
u \xrightarrow{M,V} z.
\]
The form of $u$ differs across methods. Chain-of-thought prompting makes $u$ a reasoning trace \citep{wei2022}; self-consistency makes $u$ a set of sampled reasoning paths whose agreement can be compared \citep{wang2023}; Tree of Thoughts makes $u$ a search structure of partial states \citep{yao2023a}; and related methods elicit structure by reorganizing the task or context before answering \citep{press2023,zhou2023, weston2023}. Elicitation thus supports self-governance by making reasoning, uncertainty, evidence, or task structure inspectable. 

\textbf{Reflective iteration}. Reflective iteration uses prior generation to update the path toward $m^\ast$.  An earlier output or trajectory becomes the object of later monitoring, evaluation, control, and adaptation:
\[
y_t \rightarrow u_t \xrightarrow{M,V} z_t \xrightarrow{\pi} a_t \xrightarrow{A} h_{t+1}.
\]
The reflective procedure unfolds through regulatory actions from $a_t$ to $a_{t+k}$. If a revision is selected at step $k_0$, where $k_0$ is determined to be $m^\ast$, then $a_{t+k_0}$ is a terminal action, and the revised output $y_{t+k_0}$ = $h_{t+k_0}$ marks the endpoint of the procedure, where $y$ satisfies the task goal $g$ under the metacognitive enabler and constraint $L(c,h_{t+k_0})$ within the relevant ecological context $c$. Hence, iteration contributes to effective self-governance when each pass changes the system's estimate of what remains to be done before an effective and bounded output can be reached. Otherwise, repeated generation maintains as resampling or rewriting. Self-Refine \citep{madaan2023}, Reflexion \citep{shinn2023}. Constitutional AI \citep{bai2022}. Chain-of-Verification \citep{dhuliawala2024}, retrieval-or tool-based revision\citep{gou2024,gao2023}, and dynamical system accounts of temporally evolving self-regulation \citep{ji2026,chemnitz2025} instantiate this logic in different ways. Their shared risk is that poor critique, unstable verification, or weak or problematic ecological feedback can steer the system away from $m^\ast$ rather than toward it.

\textbf{Reflective modularization}. Reflective modularization distributes the search for $m^\ast$ across the model's internal or auxiliary components such as retrievers, verifiers, memory systems, action loops, tools, and agent harnesses. Modularization maintains or changes how intermediate structure is produced, thereby where monitoring, evaluation, control, and adaptation are located:
\[
u_i = \mathcal{D}_i(x,c,h_t),
\]where $\mathcal{D}_i$ denotes a module, such as a representation, retriever, verifier, memory component, or tool interface. Modules become metacognitively meaningful only when the broader reflective procedure interprets and routes them:
\[
\{u_i\}_{i=1}^{n} \xrightarrow{M,V} z \xrightarrow{\pi} a \xrightarrow{A} h_{t+1}.
\]
Modularization helps determine $m^\ast$ by deciding when the system has enough evidence or should retrieve more evidence in retrieval-augmented generation \citep{lewis2020, zhou2024}, needs external support through tools or action interfaces \citep{schick2023, yao2023b}, should verify a claim through separate evaluative procedures \citep{dhuliawala2024, gou2024}, should store or retrieve memory for later attempts \citep{shinn2023, zhao2024}, or should escalate or stop through runtime harness control \citep{yang2024, pan2026, he2026}. This is where boundedness and effectiveness meet most directly: modules can impose or ease evidence requirements, verification gates, access rules, cost constraints, and stopping criteria, but they improve self-governance only when uncertainty, insufficiency, or failure is routed toward the right support and under the proper feedback at the ecological level, rather than committed to procedural complexity alone.

\subsection{Ecological level}

At the ecological level, the central question is when metacognitive regulation becomes meaningful in use, where context $c$ acquires its full meaning for self-governance. It specifies the conditions under which an output $y_{t+m^\ast}$ is not only task-effective, but also bounded by evidence, uncertainty, cost, authority, value, institution of its setting. 

Let
\[
\mathcal{E}_c=(I,W,Q)
\]
denote the ecological specification of context $c$, where $I$ is the \textit{reflective interface}, $W$ the \textit{reflective workflow}, and $Q$ the \textit{reflective accountability} regime. They are the conditions of uptake through which the metacognitive algorithmic procedure above obtains practical status: $I$ determines whether $z_t$ can be understood as a reason for self-governing judgment; $W$ determines whether $a_t$ can be carried into a course of self-governing action; and $Q$ determines whether $h_t \rightarrow h_{t+1}$ can be justified as an accountable self-governing transition. Together, they subject the task goal $g$ to the contextual limit $L(c,h_{t+k})$ that make the self-governing horizon $m^\ast$ ecologically proper and achievable in context \textit{c}.

We can therefore treat the algorithmic procedure as producing a \textbf{metacognitive decision event }$e_t$:
\[
e_t = (z_t, a_t, h_t \rightarrow h_{t+1}),
\]
whose practical status depends on the contextual ecology in which it occurs:
\[
\operatorname{status}(e_t \mid \mathcal{E}_c).
\]
The same event may support self-governance, be ignored, induce overreliance, create friction, or become institutionally unusable depending on $\mathcal{E}_c$. Ecology therefore determines whether metacognitive regulation can become effective and bounded in practice.

\textbf{Reflective interface}. Reflective interface $I$ concerns the ecological standing of the regulatory signal $z_t$:
\[
I: z_t \mapsto \text{\textit{legible signal}}.
\]
Metacognitive signals reach users through interface forms, and these forms shape how users trust, ignore, verify, or over-rely on AI outputs. Research on automation bias, trust calibration, and human-AI decision-making shows that their effects crucially depend on task structure, cognitive effort, verification costs, user incentives, prior beliefs, among others \citep{parasuraman1997,lee2004,lai2019,green2019,yin2019,bansal2021}. Cognitive forcing and verification-oriented interface designs can reduce overreliance by changing how users engage with AI recommendations, although they may also introduce costs in effort, usability, or workflow efficiency \citep{bucinca2021,vasconcelos2023}. Without reflective interface, $z_t$ may exist algorithmically while failing ecologically: it may be invisible, misread, over-trusted, or disconnected from what should be checked, revised, withheld, or escalated.

\textbf{Reflective workflow}. Reflective workflow $W$ concerns the ecological standing of the control action $a_t$:
\[W: a_t \mapsto \text{\textit{distributed action.}}\]
It concerns the conditions under which $a_t$  becomes embedded in a coordinated course of action across models, users, tools, documents, organizational routines, and institutional procedures. This view extends work on distributed cognition, extended cognition, and situated action, where cognitive control is organized within and across agents, artifacts, environments, and practices \citep{suchman1987,hutchins1995,clark1998,hollan2000,dourish2001}. Recent AI work makes this point increasingly concrete for generative systems by locating distributed reliability particularly in retrieval and evaluation pipelines \citep{gao2024,fan2024}, RAG failure diagnosis \citep{barnett2024}, deployed governance workflows \citep{raji2020,technology2024}.

\textbf{Reflective accountability}. Reflective accountability $Q$ concerns the governance standing of the transition from $h_t$ to $h_{t+1}$: \[Q: h_t \rightarrow h_{t+1} \mapsto \text{\textit{accountable governing}}\] Its function is to make self-regulatory procedure ultimately answerable towards proper regulatory horizon $m^\ast$: recorded, interpretable, contestable, auditable, and connected to responsible roles.  Sociotechnical AI research shows that model behavior must be evaluated in relation to deployment context, stakeholder roles, organizational incentives, and downstream harms \citep{selbst2019,suresh2021}. Practitioner-oriented work makes the same point operationally by emphasizing debugging, documentation, auditing, checklists, and accountability processes \citep{holstein2019,mitchell2019,raji2020,madaio2020,gebru2021}. Work on contestable AI further shows that accountable and responsible systems must support challenge, questioning, and redress \citep{alfrink2023}, while governance frameworks treat trustworthy AI as a contextual, organizational achievement rather than a model property alone \citep{kuehnert2025,ojewale2025}. From a metacognitive perspective, reflective governance specifies the accountability conditions under which a system's self-regulation and adaptation can guide action under unavoidable internal and environmental complexity and uncertainty \citep{grossmann2025,smith2025,ji2026,johnson2026}.

\section{Metacognitive alignment towards self-governance requires dynamic, cross-level feedback}

Metacognitive alignment requires feedback because the adequacy of a regulatory target, procedure, or contextual limit cannot be settled at one level alone. A computational target may be well specified but poorly realized; an algorithmic procedure may regulate generation but fail in use; an ecological constraint may be responsible in principle but lack the signal or action needed to make it operative. Feedback is therefore the design through which the three levels test and revise one another. 

Let 
\[
\Delta_t = F_{\mathcal{E}_c}(e_t)
\]
denote \textbf{cross-level feedback} generated when a metacognitive decision event $e_t$ enters its ecological context $c$ at time $t$ (either as updating an immediate structure $u$ or moving towards producing an output $y$). It returns to the computational and algorithmic levels:
\[
\Delta_t \Rightarrow (M,V,\pi,A,L)_{t+1}.
\]
The cross-level feedback $\Delta_t$ may identify what the regulation monitors, how signals were routed, when the procedure stopped, and whether the selected action could be interpreted, enacted, contested, or assigned to responsible roles (\hyperref[tab:mismatches]{Table~\ref*{tab:mismatches}}). In this sense, $\Delta_t$  tests both whether the ultimate $y_{t+m^\ast}$ is effective and properly bounded in ecological context. Recent work echoes this ecological point: confidence cues and system recommendations can support appropriate reliance or intensify overreliance depending on how they are presented and embedded in use \citep{parasuraman1997,lee2004,bansal2021,bucinca2021,vasconcelos2023}, and governance and contestability work makes a similar point in the institutional context \citep{raji2020,alfrink2023}. 

On the other hand, the inclusion of $L$ renders explicit the contextual conditions that both enable and constrain metacognitive regulation (see Section 3.2). Hence, feedback revises $L$ when use reveals that these conditions were mis-specified in context, thereby also reshaping the ecology in which future metacogntive decision events $e_t$ s occur. Contemporary work on human decisions and machine predictions makes this point increasingly visible: algorithmic outputs can reshape the environments in which decision-makers interpret evidence, allocate trust, act, and update knowledge \citep{kleinberg2018,ludwig2021,mullainathan2025,acemoglu2026a,acemoglu2026b}. 

In short, feedback $\Delta_t$ navigates and regulates across the three levels: ecology may revise what counts as adequate self-governance, while computational and algorithmic changes may reorganize the ecological conditions under which future metacognitive decision events $e_t$s obtain proper practical status. Together, they move towards stabilizing or changing the self-governing horizon $m^\ast$ within proper temporary, ecological context. Cross-level metacognitive alignment is therefore dynamic across the three levels, concerning what forms of agentive relations the metacognitive decision-making could achieve for bounded and effective self-governance.
\begin{table}[t]
\caption{Diagnosis of examples of cross-level metacognitive misalignment, based on the tripartite framework. The table summarizes how AI governing design may fail when metacognitive targets, procedures, and ecological conditions are misaligned.}
\label{tab:mismatches}
\centering
\scriptsize
\setlength{\tabcolsep}{2.5pt}
\renewcommand{\arraystretch}{1.15}
\begin{tabularx}{\textwidth}{p{0.19\textwidth} p{0.12\textwidth} p{0.12\textwidth} X p{0.18\textwidth}}
\toprule
\textbf{Potential cross-level mismatch} & \textbf{Strength} & \textbf{Weakness} & \textbf{Example} & \textbf{What the framework can add} \\
\midrule
Computationally vague, algorithmically elaborate & Reflective procedure & Meta-level function & Chain-of-thought elicitation \citep{wei2022}; rationale unfaithfulness \citep{turpin2023}; generic reflection prompting \citep{wang2024} & Metacognitive functional specification \\
Computationally valid, algorithmically weak & Valid target & Reliable signal & Verbalized uncertainty \citep{lin2022,xiong2024}; self-knowledge limits \citep{kadavath2022}; confidence calibration instability \citep{tian2023}; generation-verification mismatch \citep{song2024} & Metacognitive signal validation \\
Algorithmically strong, ecologically brittle & Effective procedure & Situated use & Trust calibration failure \citep{lee2004}; human-AI complementarity breakdown \citep{bansal2021}; cognitive forcing need \citep{bucinca2021}; overreliance under AI advice \citep{vasconcelos2023} & Use-context reflectivity \\
Ecologically responsible, computationally underspecified & Governance aim & Triggering signal & Ethics checklist without runtime criteria \citep{madaio2020}; audit process without escalation signals \citep{raji2020}; documentation without intervention thresholds \citep{mitchell2019,gebru2021}; risk workflow without operational monitors \citep{technology2024} & Metacognitively triggered signal-to-action alignment \\
\bottomrule
\end{tabularx}
\end{table}

\section{A Case Study}
The case of algorithmic recourse makes the cross-level metacognitive alignment for self-governance concrete. Consider a loan applicant denied by an automated classifier. A counterfactual explanation may identify a model-flipping change, such as increasing income by a specified amount \citep{wachter2017, wachter2018}. Actionable and causal recourse then ask whether such a change is feasible, interventionally warranted, and connected to a real path from the applicant's current state to the desired outcome \citep{ustun2019, karimi2020, karimi2021}. Yet even a technically valid recommendation is not self-authorizing: it may depend on uncertain causal knowledge, unequal access to remedy, model or distributional change, institutional contingency, and sociocultural recognition, even when the system provisionally treats its evaluation as fair or justified.

In this setting, the metacognitive decision event $e_t = (z_t,a_t,h_t \rightarrow h_{t+1})$ is the point at which the system assesses a candidate recourse recommendation and decides whether it should be presented, qualified, revised, routed, or withheld. Here $z_t$ summarizes the recommendation's regulatory status, including its feasibility, causal warrant, uncertainty under various agentive contexts \citep{khotanlou2025}, while $a_t$ selects and performs the appropriate treatment of that recommendation. Cross-level metacognitive feedback $\Delta_t = F_{E_c}(e_t)$ then diagnoses potential cross-level misalignment: a recommendation may satisfy the classifier but fail as ecologically actionable recourse, appear procedurally fair while lacking causal warrant, or be causally coherent while lacking institutional uptake or contestable status. In such cases, $\Delta_t$ should revise not only the recommendation itself, but also the metacognitive constraint $L(c,h_{t+k})$ under which future recourse outputs are treated as warranted, usable, and accountable. The self-governing horizon $m^\ast$ is reached only when the system has produced a recourse output $y_{t+m^\ast}$ that is sufficient for the object-level goal, helping the applicant understand, contest, or change the decision, while remaining bounded by the metacognitive constraints $L(c,h_{t+k})$ determined by both the algorithm and ecology. Those limits concern not merely causal assumptions or cost as task-level features, but the system's authority to present advice as warranted, actionable, fair, and institutionally usable. 

This case therefore clarifies bounded and effective self-governance: recourse is effective when it improves an agent's practical ability to act on or challenge a decision, and bounded when the system regulates the authority, uncertainty, and ecological contingency of its own guidance. Algorithmic recourse thus shows why metacognition should not be an extra, post-hoc layer after generation and prediction. It is the self-governing process through which a system decides why a recommendation is warranted, how it should be qualified, and under what ecological conditions it can become responsibly actionable.

\section{Research Agenda}

A research agenda for bounded and effective self-governance in generative AI should move in four directions. The first is to \textbf{treat capability and governance as mutually enabling }rather than competing aims. Generative systems should and can be evaluated by whether their competence includes the capacity to regulate and govern their own exercise. The second is \textbf{humble, perspectival self-governance} against a purely maximalist view of AI capability \citep{johnson2026}. The field should not define advanced AI only by ever greater task-level behavioral reach, but also by the ability to recognize the authority, uncertainty, and perspective-limits of its own outputs, and to revise, qualify, or defer accordingly. The first and second directions together require the development of new datasets and benchmarks that test \textit{both} task performance \textit{and} whether systems can manage and regulate the exercise of that performance. The third is \textbf{cross-level diagnosis through dynamic feedback}. Metacognitive goals and procedures should be studied as self-governing pathways whose failures may become visible only when use feeds back on what was regulated, how it was regulated, and whether the contextual limits were properly specified (\hyperref[tab:mismatches]{Table~\ref*{tab:mismatches}}). The fourth is \textbf{distributed and higher-order metacognition}. As self-governance increasingly depends on external and distributed supports, the capacity to decide why, how, and when to rely on those supports should be treated as intrinsic to self-governance itself. In particular, self-governing systems must be able not only to regulate task- level activity in the first-order manner, but also to manage their own regulatory behavior through higher-order metacognitive control \citep{langlois2020}.

\section{Alternative Views and Objections}

Our position is sharpened by distinguishing it from three narrower views of metacognition. First, \textbf{metacognition may seem to anthropomorphize AI by importing a human mental model}. Although the term is inspired by human cognition, it does not locate metacognition inside one person's mind or one model's inner state. We use metacognition functionally: It concerns whether information about a system's own states, outputs, and trajectories is organized to regulate later generative behavior, and the locus of such regulation may be internal, scaffolded, or distributed across models, tools, users, organizations, and so on, consistent with computational accounts of metacognition and metareasoning as control over a system's own reasoning processes \citep{russell1991,cox2011}.

Second, \textbf{metacognition may seem to mean deliberative, expensive, and ``slow'' rationality}. This view appears in parts of metacognition and System-2-inspired AI work \citep{wei2022,madaan2023,dhuliawala2024}, often through psychology-inspired dual-process language \citep{kahneman2011,evans2013}. But deliberation does not always improve judgment or performance, as work on bounded and ecological rationality has long emphasized \citep{gigerenzer1996,kruglanski2011,gigerenzer2025a,gigerenzer2025b}. Metacognition is therefore not simply more resource-intensive reflection for better outputs \citep{ji2026,katsikopoulos2026}, and can support resource-saving regulation when the appropriate self-governing horizon is determined in certain contexts.

Third, \textbf{distributed and higher-order metacognition may seem to make self-governance too expansive and costly}. Once metacognition extends to interfaces, workflows, and regulatory supports, self-governance may seem unlimitedly open-ended. But distributed and higher-order metacognition is not unconstrained regulation. In particular, higher-order, meta- metacognition is one crucial capacity to bound regulation itself \citep{langlois2020}. Distributed and higher-order metacognition therefore strengthens rather than weakens the position: it explains how self-governance can remain effective without becoming either isolated inside the model or expanded without limit.

\section{Conclusion}
This paper has argued that generative AI needs metacognition to become bounded and effectively self-governing. By specifying metacognition across computational targets, algorithmic procedures, and ecological arrangements, the account developed here makes self-governance a concrete object of design and evaluation. Metacognition thereby makes it possible to conceive generative AI as both capable and well-governed, rather than treating capability and governance as competing aims.

\bibliographystyle{plainnat}
\bibliography{references}

@misc{acemoglu2026a,
  author = {Daron Acemoglu and Dingwen Kong and Asuman Ozdaglar},
  title = {{AI, Human Cognition and Knowledge Collapse}},
  year = {2026},
  doi = {10.3386/w34910},
  note = {NBER Working Paper No. 34910. National Bureau of Economic Research}
}

@misc{acemoglu2026b,
  author = {Daron Acemoglu and Tianyi Lin and Asuman Ozdaglar and James Siderius},
  title = {{How AI Aggregation Affects Knowledge. NBER Working Paper No. 35036}},
  year = {2026},
  doi = {10.3386/w35036},
  note = {National Bureau of Economic Research}
}

@misc{alfrink2023,
  author = {Kars Alfrink and Ianus Keller and Neelke Doorn and Gerd Kortuem},
  title = {{Contestable AI by Design: Towards a Framework}},
  year = {2023},
  doi = {10.1007/s11023-023-09611-z},
  note = {Minds and Machines 33: 613-639}
}

@misc{azaria2023,
  author = {Amos Azaria and Tom M Mitchell},
  title = {{The Internal State of an LLM Knows When Its Lying}},
  year = {2023},
  note = {arXiv:2304.13734.}
}

@misc{bai2022,
  author = {Yuntao Bai and Saurav Kadavath and Sandipan Kundu and Amanda Askell and Jackson Kernion and Andy Jones and others},
  title = {{Constitutional AI: Harmlessness from AI Feedback}},
  year = {2022},
  note = {arXiv:2212.08073.}
}

@misc{bansal2021,
  author = {Gagan Bansal and Tongshuang Wu and Joyce Zhou and Raymond Fok and Besmira Nushi and Ece Kamar and Marco Tulio Ribeiro and Daniel S. Weld},
  title = {{Does the Whole Exceed Its Parts? The Effect of AI Explanations on Complementary Team Performance}},
  year = {2021},
  doi = {10.1145/3411764.3445717},
  note = {In Proceedings of the 2021 CHI Conference on Human Factors in Computing Systems, 1--16. New York: Association for Computing Machinery}
}

@misc{barnett2024,
  author = {Scott Barnett and Stefanus Kurniawan and Srikanth Thudumu and Zach Brannelly and Mohamed Abdelrazek},
  title = {{Seven Failure Points When Engineering a Retrieval Augmented Generation System}},
  year = {2024},
  note = {arXiv:2401.05856.}
}

@misc{bechtel2015,
  author = {William Bechtel and Oron Shagrir},
  title = {{The Non-Redundant Contributions of Marr's Three Levels of Analysis for Explaining Information-Processing Mechanisms}},
  year = {2015},
  doi = {10.1111/tops.12141},
  note = {Topics in Cognitive Science 7(2): 312-322}
}

@misc{bucinca2021,
  author = {Zana Buçinca and Maja Barbara Malaya and Krzysztof Z. Gajos},
  title = {{To Trust or to Think: Cognitive Forcing Functions Can Reduce Overreliance on AI in AI-Assisted Decision-Making}},
  year = {2021},
  doi = {10.1145/3449287},
  note = {Proceedings of the ACM on Human-Computer Interaction 5(CSCW1): Article 188, 1--21}
}

@misc{caro2022,
  author = {Manuel F. Caro and Michael T. Cox and Raúl E. Toscano-Miranda},
  title = {{A Validated Ontology for Metareasoning in Intelligent Systems}},
  year = {2022},
  doi = {10.3390/jintelligence10040113},
  note = {Journal of Intelligence 10(4): 113}
}

@misc{chemnitz2025,
  author = {Dennis Chemnitz and Maximilian Engel and Christian Kuehn and Sara-Viola Kuntz},
  title = {{A Dynamical Systems Perspective on the Analysis of Neural Networks}},
  year = {2025},
  note = {arXiv:2507.05164.}
}

@misc{clark1998,
  author = {Andy Clark and David J Chalmers},
  title = {{The Extended Mind}},
  year = {1998},
  doi = {10.1093/analys/58.1.7},
  note = {Analysis 58(1): 7-19}
}

@misc{cox2005,
  author = {Michael T Cox},
  title = {{Metacognition in Computation: A Selected Research Review}},
  year = {2005},
  doi = {10.1016/j.artint},
  note = {Artificial Intelligence 169(2): 104-141.  2005.10.009}
}

@misc{cox2011,
  author = {Michael T. Cox and Anita Raja},
  title = {{Metareasoning: Thinking about Thinking}},
  year = {2011},
  note = {Cambridge, MA: MIT Press.}
}

@misc{cox2022,
  author = {Michael T. Cox and Anita Raja and Dustin Dannenhauer and Matthew Molineaux and Hector Munoz-Avila},
  title = {{Computational Metacognition}},
  year = {2022},
  doi = {10.1002/aaai.12050},
  note = {AI Magazine 43(2): 192-203}
}

@misc{dhuliawala2024,
  author = {Shehzaad Dhuliawala and Mojtaba Komeili and Jing Xu and Roberta Raileanu and Xian Li and Asli Celikyilmaz and others},
  title = {{Chain-of-Verification Reduces Hallucination in Large Language Models}},
  year = {2024},
  doi = {10.18653/v1/2024.findings-acl.212},
  note = {In Findings of the Association for Computational Linguistics: ACL 2024. Association for Computational Linguistics, pages 3563-3578}
}

@misc{dourish2001,
  author = {Paul Dourish},
  title = {{Where the Action Is: The Foundations of Embodied Interaction}},
  year = {2001},
  note = {MIT Press.}
}

@misc{dunlosky2009,
  author = {John Dunlosky and Janet Metcalfe},
  title = {{Metacognition}},
  year = {2009},
  note = {SAGE Publications.}
}

@misc{efklides2008,
  author = {Anastasia Efklides},
  title = {{Metacognition: Defining Its Facets and Levels of Functioning in Relation to Self-Regulation and Co-Regulation}},
  year = {2008},
  doi = {10.1027/1016-9040.13.4.277},
  note = {European Psychologist 13(4): 277-287}
}

@misc{efklides2011,
  author = {Anastasia Efklides},
  title = {{Interactions of Metacognition with Motivation and Affect in Self-Regulated Learning: The MASRL Model}},
  year = {2011},
  doi = {10.1080/00461520},
  note = {Educational Psychologist 46(1): 6-25.  2011.538645}
}

@misc{evans2013,
  author = {Jonathan St. B. T. Evans and Keith E Stanovich},
  title = {{Dual-Process Theories of Higher Cognition: Advancing the Debate}},
  year = {2013},
  doi = {10.1177/1745691612460685},
  note = {Perspectives on Psychological Science 8(3): 223-241}
}

@misc{fan2024,
  author = {Wenqi Fan and Yujuan Ding and Liangbo Ning and Shijie Wang and Hengyuan Li and Dawei Yin and others},
  title = {{A Survey on RAG Meeting LLMs: Towards Retrieval-Augmented Large Language Models}},
  year = {2024},
  doi = {10.1145/3637528.3671470},
  note = {In Proceedings of the 30th ACM SIGKDD Conference on Knowledge Discovery and Data Mining. ACM, pages 6491-6501}
}

@misc{flavell1979,
  author = {John H Flavell},
  title = {{Metacognition and Cognitive Monitoring: A New Area of Cognitive--Developmental Inquiry}},
  year = {1979},
  doi = {10.1037/0003-066X.34.10.906},
  note = {American Psychologist 34(10): 906-911}
}

@misc{fleming2012,
  author = {Stephen M. Fleming and Raymond J Dolan},
  title = {{The Neural Basis of Metacognitive Ability}},
  year = {2012},
  doi = {10.1098/rstb},
  note = {Philosophical Transactions of the Royal Society B: Biological Sciences 367(1594): 1338-1349.  2011.0417}
}

@misc{gao2023,
  author = {Luyu Gao and Zhuyun Dai and Panupong Pasupat and Anthony Chen and Arun Tejasvi Chaganty and Angela Fan and others},
  title = {{RARR: Researching and Revising What Language Models Say, Using Language Models}},
  year = {2023},
  doi = {10.18653/v1/2023.acl-long.910},
  note = {In Proceedings of the 61st Annual Meeting of the Association for Computational Linguistics, 16477--16508. Association for Computational Linguistics}
}

@misc{gao2024,
  author = {Yunfan Gao and Yun Xiong and Xinyu Gao and Kangxiang Jia and Jinliu Pan and Yuxi Bi and others},
  title = {{Retrieval-Augmented Generation for Large Language Models: A Survey}},
  year = {2024},
  note = {arXiv:2312.10997.}
}

@misc{gebru2021,
  author = {Timnit Gebru and Jamie Morgenstern and Briana Vecchione and Jennifer Wortman Vaughan and Hanna Wallach and Hal Daumé and others},
  title = {{Datasheets for Datasets}},
  year = {2021},
  doi = {10.1145/3458723},
  note = {Communications of the ACM 64(12): 86-92}
}

@misc{gigerenzer2025a,
  author = {Gerd Gigerenzer},
  title = {{The Legacy of Daniel Kahneman: A Personal View. Erasmus Journal for Philosophy and Economics 18(1): 28--61}},
  year = {2025},
  doi = {10.23941/ejpe.v18i1.1075}
}

@misc{gigerenzer2025b,
  author = {Gerd Gigerenzer},
  title = {{Ecological Rationality: Rethinking Behavioural Economics}},
  year = {2025},
  doi = {10.4337/9781802207736.00052},
  note = {In Elgar Encyclopedia of Behavioural and Experimental Economics, 153--155. Edward Elgar}
}

@misc{gigerenzer1996,
  author = {Gerd Gigerenzer and Daniel G. Goldstein},
  title = {{Reasoning the Fast and Frugal Way: Models of Bounded Rationality}},
  year = {1996},
  doi = {10.1037/0033-295X.103.4.650},
  note = {Psychological Review 103(4): 650--669}
}

@misc{gou2024,
  author = {Zhibin Gou and Zhihong Shao and Yeyun Gong and Yelong Shen and Yujiu Yang and Nan Duan and others},
  title = {{CRITIC: Large Language Models Can Self-Correct with Tool-Interactive Critiquing}},
  year = {2024},
  note = {In Proceedings of the Twelfth International Conference on Learning Representations.}
}

@misc{green2019,
  author = {Ben Green and Yiling Chen},
  title = {{The Principles and Limits of Algorithm-in-the-Loop Decision Making}},
  year = {2019},
  doi = {10.1145/3359152},
  note = {Proceedings of the ACM on Human-Computer Interaction 3(CSCW): Article 50, 1--24}
}

@misc{griffiths2015,
  author = {Thomas L. Griffiths and Falk Lieder and Noah D Goodman},
  title = {{Rational Use of Cognitive Resources: Levels of Analysis between the Computational and the Algorithmic}},
  year = {2015},
  doi = {10.1111/tops.12142},
  note = {Topics in Cognitive Science 7(2): 217-229}
}

@misc{grossmann2017,
  author = {Igor Grossmann},
  title = {{Wisdom in Context}},
  year = {2017},
  doi = {10.1177/1745691616672066},
  note = {Perspectives on Psychological Science 12(2): 233-257}
}

@misc{grossmann2021,
  author = {Igor Grossmann and Harrison Oakes and Henri C Santos},
  title = {{Training for Wisdom: The Distanced-Self-Reflection Diary Method}},
  year = {2021},
  doi = {10.1177/0956797620969170},
  note = {Psychological Science 32(3): 381-394}
}

@misc{grossmann2020,
  author = {Igor Grossmann and Nic M. Weststrate and Monika Ardelt and Justin P. Brienza and Mengxi Dong and Michel Ferrari and others},
  title = {{The Science of Wisdom in a Polarized World: Knowns and Unknowns}},
  year = {2020},
  doi = {10.1080/1047840X},
  note = {Psychological Inquiry 31(2): 103-133.  2020.1750917}
}

@misc{grossmann2025,
  author = {Igor Grossmann and Samuel G. B. Johnson},
  title = {{Cultivating Wisdom Through Metacognition: A New Frontier in Decision-Making Under Radical Uncertainty}},
  year = {2025},
  doi = {10.1037/mac0000235},
  note = {Journal of Applied Research in Memory and Cognition 14(3): 301--318}
}

@misc{hamrick2020,
  author = {Jessica B. Hamrick and Shakir Mohamed},
  title = {{Levels of Analysis for Machine Learning}},
  year = {2020},
  note = {arXiv:2004.05107.}
}

@misc{he2026,
  author = {Chaoyue He and Xin Zhou and Di Wang and Hong Xu and Wei Liu and Chunyan Miao},
  title = {{Harness Engineering for Language Agents: The Harness Layer as Control, Agency, and Runtime}},
  year = {2026},
  doi = {10.20944/preprints202603.1756.v2},
  note = {Preprints, version 2, April 2026}
}

@misc{hollan2000,
  author = {James Hollan and Edwin Hutchins and David Kirsh},
  title = {{Distributed Cognition: Toward a New Foundation for Human--Computer Interaction Research}},
  year = {2000},
  doi = {10.1145/353485.353487},
  note = {ACM Transactions on Computer-Human Interaction 7(2): 174-196}
}

@misc{holstein2019,
  author = {Kenneth Holstein and Jennifer Wortman Vaughan and Hal Daumé III and Miro Dudík and Hanna Wallach},
  title = {{Improving Fairness in Machine Learning Systems: What Do Industry Practitioners Need? In Proceedings of the 2019 CHI Conference on Human Factors in Computing Systems}},
  year = {2019},
  doi = {10.1145/3290605.3300830},
  note = {ACM}
}

@misc{huang2024,
  author = {Jie Huang and Xinyun Chen and Swaroop Mishra and Huaixiu Steven Zheng and Adams Wei Yu and Xinying Song and others},
  title = {{Large Language Models Cannot Self-Correct Reasoning Yet}},
  year = {2024},
  note = {In Proceedings of the Twelfth International Conference on Learning Representations.}
}

@misc{hutchins1995,
  author = {Edwin Hutchins},
  title = {{Cognition in the Wild}},
  year = {1995},
  note = {MIT Press.}
}

@misc{ji2026,
  author = {Eugene Yu Ji},
  title = {{Unraveling Authoritarian Reform Strategies: A Metacognitive-Subcognitive Model}},
  year = {2026},
  doi = {10.1111/ajps.70054},
  note = {American Journal of Political Science, 1--20}
}

@misc{jian2025,
  author = {Li Ji-An and Hua-Dong Xiong and Robert C. Wilson and Marcelo G. Mattar and Marcus K Benna},
  title = {{Language Models Are Capable of Metacognitive Monitoring and Control of Their Internal Activations}},
  year = {2025},
  note = {arXiv:2505.13763.}
}

@misc{johnson2026,
  author = {Samuel G. B. Johnson and Amir-Hossein Karimi and Yoshua Bengio and Nick Chater and Tobias Gerstenberg and Kate Larson and  Kate Larson and Sydney Levine and Melanie Mitchell and Iyad Rahwan and Bernhard Schölkopf and Igor Grossmann},
  title = {{Imagining and Building Wise Machines: The Centrality of AI Metacognition}},
  year = {2026},
  doi = {10.1016/j.tics},
  note = {Trends in Cognitive Sciences.  2026.01.002}
}

@misc{kadavath2022,
  author = {Saurav Kadavath and Tom Conerly and Amanda Askell and Tom Henighan and Dawn Drain and Ethan Perez and others},
  title = {{Language Models (Mostly) Know What They Know}},
  year = {2022},
  note = {arXiv:2207.05221.}
}

@misc{kahneman2011,
  author = {Daniel Kahneman},
  title = {{Thinking, Fast and Slow}},
  year = {2011},
  note = {Farrar, Straus and Giroux.}
}

@misc{kapoor2024,
  author = {Sanyam Kapoor and Nate Gruver and Manley Roberts and Katherine Collins and Arka Pal and Umang Bhatt and others},
  title = {{Large Language Models Must Be Taught to Know What They Don't Know}},
  year = {2024},
  note = {In Advances in Neural Information Processing Systems.}
}

@misc{katsikopoulos2026,
  author = {Konstantinos V. Katsikopoulos and Gerd Gigerenzer},
  title = {{Fast-and-Frugal Heuristics: Analytical Models of Intuition. IMA Journal of Management Mathematics 37(1): 17--33}},
  year = {2026},
  doi = {10.1093/imaman/dpaf041}
}

@misc{karimi2020,
  author = {Amir-Hossein Karimi and Julius von Kügelgen and Bernhard Schölkopf and Isabel Valera},
  title = {{Algorithmic Recourse under Imperfect Causal Knowledge: A Probabilistic Approach}},
  year = {2020},
  note = {In Advances in Neural Information Processing Systems 33, 265--277.}
}

@misc{karimi2021,
  author = {Amir-Hossein Karimi and Gilles Barthe and Bernhard Schölkopf and Isabel Valera},
  title = {{Algorithmic Recourse: From Counterfactual Explanations to Interventions}},
  year = {2021},
  doi = {10.1145/3442188.3445899},
  note = {In Proceedings of the 2021 ACM Conference on Fairness, Accountability, and Transparency. ACM, pages 353-362}
}

@misc{khotanlou2025,
  author = {Zahra Khotanlou and Kate Larson and Amir-Hossein Karimi},
  title = {{From Individual to Multi-Agent Algorithmic Recourse: Minimizing the Welfare Gap via Capacitated Bipartite Matching}},
  year = {2025},
  note = {arXiv:2508.11070.}
}

@misc{kleinberg2018,
  author = {Jon Kleinberg and Himabindu Lakkaraju and Jure Leskovec and Jens Ludwig and Sendhil Mullainathan},
  title = {{Human Decisions and Machine Predictions}},
  year = {2018},
  doi = {10.1093/qje/qjx032},
  note = {The Quarterly Journal of Economics 133 (1): 237--293}
}

@misc{krafft2018,
  author = {Peter M. Krafft and Thomas L. Griffiths},
  title = {{Levels of Analysis in Computational Social Science}},
  year = {2018},
  note = {In Proceedings of the 40th Annual Conference of the Cognitive Science Society, 1963--1968. Cognitive Science Society.}
}

@misc{kruglanski2011,
  author = {Arie W. Kruglanski and Gerd Gigerenzer},
  title = {{Intuitive and Deliberate Judgments Are Based on Common Principles}},
  year = {2011},
  doi = {10.1037/a0020762},
  note = {Psychological Review 118(1): 97-109}
}

@misc{ku2025,
  author = {Alexander Y. Ku and Declan Campbell and Xuechunzi Bai and Jiayi Geng and Ryan Liu and Raja Marjieh and R. Thomas McCoy and Andrew Nam and Ilia Sucholutsky and Veniamin Veselovsky and Liyi Zhang and Jian-Qiao Zhu and Thomas L. Griffiths},
  title = {{Levels of Analysis for Large Language Models}},
  year = {2025},
  note = {arXiv:2503.13401.}
}

@misc{kuehnert2025,
  author = {Blaine Kuehnert and Rachel M. Kim and Jodi Forlizzi and Hoda Heidari},
  title = {{The 'Who,' 'What,' and 'How' of Responsible AI Governance: A Systematic Review and Meta-Analysis of (Actor, Stage)-Specific Tools}},
  year = {2025},
  doi = {10.1145/3715275.3732191},
  note = {In Proceedings of the 2025 ACM Conference on Fairness, Accountability, and Transparency, 2991--3005. New York: Association for Computing Machinery}
}

@misc{kuhn2023,
  author = {Lorenz Kuhn and Yarin Gal and Sebastian Farquhar},
  title = {{Semantic Uncertainty: Linguistic Invariances for Uncertainty Estimation in Natural Language Generation}},
  year = {2023},
  note = {In Proceedings of the Eleventh International Conference on Learning Representations.}
}

@misc{lai2019,
  author = {Vivian Lai and Chenhao Tan},
  title = {{On Human Predictions with Explanations and Predictions of Machine Learning Models: A Case Study on Deception Detection}},
  year = {2019},
  doi = {10.1145/3287560.3287590},
  note = {In Proceedings of the Conference on Fairness, Accountability, and Transparency. ACM, pages 29-38}
}

@misc{langlois2020,
  author = {Steven T. Langlois and Omotara Akoroda and Esmeralda Carrillo and Jeffrey W. Herrmann and Shapour Azarm and Huan Xu and Michael Otte},
  title = {{Metareasoning Structures, Problems, and Modes for Multiagent Systems: A Survey}},
  year = {2020},
  doi = {10.1109/ACCESS.2020.3028751},
  note = {IEEE Access 8: 183080--183089}
}

@misc{lee2004,
  author = {John D. Lee and Katrina A See},
  title = {{Trust in Automation: Designing for Appropriate Reliance}},
  year = {2004},
  doi = {10.1518/hfes.46.1.50\_30392},
  note = {Human Factors 46(1): 50-80}
}

@misc{lewis2020,
  author = {Patrick Lewis and Ethan Perez and Aleksandra Piktus and Fabio Petroni and Vladimir Karpukhin and Naman Goyal and others},
  title = {{Retrieval-Augmented Generation for Knowledge-Intensive NLP Tasks}},
  year = {2020},
  note = {In Advances in Neural Information Processing Systems 33, 9459--9474.}
}

@misc{li2024,
  author = {Loka Li and Guangyi Chen and Yusheng Su and Zhenhao Chen and Yixuan Zhang and Eric P. Xing and others},
  title = {{Confidence Matters: Revisiting Intrinsic Self-Correction Capabilities of Large Language Models}},
  year = {2024},
  note = {arXiv:2402.12563.}
}

@misc{lin2022,
  author = {Stephanie Lin and Jacob Hilton and Owain Evans},
  title = {{Teaching Models to Express Their Uncertainty in Words}},
  year = {2022},
  note = {Transactions on Machine Learning Research.}
}

@misc{liu2025,
  author = {Tennison Liu and Mihaela van der Schaar},
  title = {{Position: Truly Self-Improving Agents Require Intrinsic Metacognitive Learning}},
  year = {2025},
  note = {In Proceedings of the 42nd International Conference on Machine Learning, PMLR 267:81714--81727.}
}

@misc{ludwig2021,
  author = {Jens Ludwig and Sendhil Mullainathan},
  title = {{Fragile Algorithms and Fallible Decision-Makers: Lessons from the Justice System. Journal of Economic Perspectives 35 (4): 71--96}},
  year = {2021},
  doi = {10.1257/jep.35.4.71}
}

@misc{madaan2023,
  author = {Aman Madaan and Niket Tandon and Prakhar Gupta and Skyler Hallinan and Luyu Gao and Sarah Wiegreffe and others},
  title = {{Self-Refine: Iterative Refinement with Self-Feedback}},
  year = {2023},
  doi = {10.52202/075280-2019},
  note = {In Advances in Neural Information Processing Systems 36, 46534--46594}
}

@misc{madaio2020,
  author = {Michael A. Madaio and Luke Stark and Jennifer Wortman Vaughan and Hanna Wallach},
  title = {{Co-Designing Checklists to Understand Organizational Challenges and Opportunities around Fairness in AI}},
  year = {2020},
  doi = {10.1145/3313831.3376445},
  note = {In Proceedings of the 2020 CHI Conference on Human Factors in Computing Systems, 1--14. New York: Association for Computing Machinery}
}

@misc{marr1982,
  author = {David Marr},
  title = {{Vision: A Computational Investigation into the Human Representation and Processing of Visual Information}},
  year = {1982},
  note = {W. H. Freeman.}
}

@misc{metcalfe1994,
  author = {Janet Metcalfe and Arthur P Shimamura},
  title = {{Metacognition: Knowing about Knowing}},
  year = {1994},
  note = {MIT Press.}
}

@misc{mitchell2019,
  author = {Margaret Mitchell and Simone Wu and Andrew Zaldivar and Parker Barnes and Lucy Vasserman and Ben Hutchinson and others},
  title = {{Model Cards for Model Reporting}},
  year = {2019},
  doi = {10.1145/3287560.3287596},
  note = {In Proceedings of the Conference on Fairness, Accountability, and Transparency. ACM, pages 220-229}
}

@misc{mullainathan2025,
  author = {Sendhil Mullainathan},
  title = {{Economics in the Age of Algorithms. AEA Papers and Proceedings 115: 1--23}},
  year = {2025},
  note = {https://doi.org/10.1257/pandp. 20251118.}
}

@misc{technology2024,
  author = {{{National Institute of Standards and Technology}}},
  title = {{Artificial Intelligence Risk Management Framework: Generative Artificial Intelligence Profile}},
  year = {2024},
  doi = {10.6028/NIST.AI.600-1},
  note = {National Institute of Standards and Technology}
}

@misc{nelson1990,
  author = {Thomas O. Nelson and Louis Narens},
  title = {{Metamemory: A Theoretical Framework and New Findings}},
  year = {1990},
  doi = {10.1016/S0079-7421(08)60053-5},
  note = {In The Psychology of Learning and Motivation. Academic Press, pages 125-173}
}

@misc{ojewale2025,
  author = {Victor Ojewale and Ryan Steed and Briana Vecchione and Abeba Birhane and Inioluwa Deborah Raji},
  title = {{Towards AI Accountability Infrastructure: Gaps and Opportunities in AI Audit Tooling}},
  year = {2025},
  doi = {10.1145/3706598.3713301},
  note = {In Proceedings of the 2025 CHI Conference on Human Factors in Computing Systems, Article 815, 1--29. New York: Association for Computing Machinery}
}

@misc{overgaard2012,
  author = {Morten Overgaard and Kristian Sandberg},
  title = {{Kinds of Access: Different Methods for Report Reveal Different Kinds of Metacognitive Access}},
  year = {2012},
  doi = {10.1098/rstb},
  note = {Philosophical Transactions of the Royal Society B: Biological Sciences 367(1594): 1287-1296.  2011.0425}
}

@misc{pan2026,
  author = {Linyue Pan and Lexiao Zou and Shuo Guo and Jingchen Ni and Hai-Tao Zheng},
  title = {{Natural-Language Agent Harnesses}},
  year = {2026},
  note = {arXiv:2603.25723.}
}

@misc{parasuraman1997,
  author = {Raja Parasuraman and Victor Riley},
  title = {{Humans and Automation: Use, Misuse, Disuse, Abuse}},
  year = {1997},
  doi = {10.1518/001872097778543886},
  note = {Human Factors 39(2): 230-253}
}

@misc{press2023,
  author = {Ofir Press and Muru Zhang and Sewon Min and Ludwig Schmidt and Noah A. Smith and Mike Lewis},
  title = {{Measuring and Narrowing the Compositionality Gap in Language Models}},
  year = {2023},
  doi = {10.18653/v1/2023.findings-emnlp.378},
  note = {In Findings of the Association for Computational Linguistics: EMNLP 2023, 5687--5711. Association for Computational Linguistics}
}

@misc{proust2013,
  author = {Joelle Proust},
  title = {{The Philosophy of Metacognition: Mental Agency and Self-Awareness}},
  year = {2013},
  note = {Oxford University Press.}
}

@misc{raji2020,
  author = {Inioluwa Deborah Raji and Andrew Smart and Rebecca N. White and Margaret Mitchell and Timnit Gebru and Ben Hutchinson and others},
  title = {{Closing the AI Accountability Gap: Defining an End-to-End Framework for Internal Algorithmic Auditing}},
  year = {2020},
  doi = {10.1145/3351095.3372873},
  note = {In Proceedings of the 2020 Conference on Fairness, Accountability, and Transparency. ACM, pages 33-44}
}

@misc{ren2023,
  author = {Jie Ren and Yao Zhao and Tu Vu and Peter J. Liu and Balaji Lakshminarayanan},
  title = {{Self-Evaluation Improves Selective Generation in Large Language Models}},
  year = {2023},
  note = {In Proceedings on I Can't Believe It's Not Better: Failure Modes in the Age of Foundation Models at NeurIPS 2023 Workshops, PMLR 239:49--64.}
}

@misc{russell1989,
  author = {Stuart J. Russell and Eric H Wefald},
  title = {{Principles of Metareasoning}},
  year = {1989},
  note = {In Proceedings of the First International Conference on Principles of Knowledge Representation and Reasoning. Morgan Kaufmann, pages 400-411.}
}

@misc{russell1991,
  author = {Stuart J. Russell and Eric H Wefald},
  title = {{Do the Right Thing: Studies in Limited Rationality}},
  year = {1991},
  note = {MIT Press.}
}

@misc{schick2023,
  author = {Timo Schick and Jane Dwivedi-Yu and Roberto Dessì and Roberta Raileanu and Maria Lomeli and Eric Hambro and Luke Zettlemoyer and Nicola Cancedda and Thomas Scialom},
  title = {{Toolformer: Language Models Can Teach Themselves to Use Tools}},
  year = {2023},
  note = {In Advances in Neural Information Processing Systems 36, 68539--68551.}
}

@misc{schmill2011,
  author = {Matthew D. Schmill and Michael L. Anderson and Scott Fults and Darsana Josyula and Tim Oates and Donald Perlis and Hamid Haidarian Shahri and Shomir Wilson and Dean Wright},
  title = {{The Metacognitive Loop and Reasoning about Anomalies}},
  year = {2011},
  note = {In Metareasoning: Thinking about Thinking, edited by Michael T. Cox and Anita Raja, 183--198. Cambridge, MA: MIT Press.}
}

@misc{selbst2019,
  author = {Andrew D. Selbst and Danah Boyd and Sorelle A. Friedler and Suresh Venkatasubramanian and Janet Vertesi},
  title = {{Fairness and Abstraction in Sociotechnical Systems}},
  year = {2019},
  doi = {10.1145/3287560.3287598},
  note = {In Proceedings of the Conference on Fairness, Accountability, and Transparency. ACM, pages 59-68}
}

@misc{servajean2026,
  author = {Richard Servajean and Philippe Servajean},
  title = {{Measuring the Metacognition of AI}},
  year = {2026},
  note = {arXiv:2603.29693. arXiv:2603.29693.}
}

@misc{shinn2023,
  author = {Noah Shinn and Federico Cassano and Ashwin Gopinath and Karthik R. Narasimhan and Shunyu Yao},
  title = {{Reflexion: Language Agents with Verbal Reinforcement Learning}},
  year = {2023},
  note = {In Advances in Neural Information Processing Systems 36, 8634--8652.}
}

@misc{smith2025,
  author = {Jackson A. Smith and Anna Dorfman and Neil Wegenschimmel and Igor Grossmann},
  title = {{Wisdom Reconsidered: A Dynamic Network Account of Metacognition and Complex Thought}},
  year = {2025},
  doi = {10.1037/xge0001821},
  note = {Journal of Experimental Psychology: General 154(11): 3114--3128}
}

@misc{song2024,
  author = {Yuda Song and Hanlin Zhang and Carson Eisenach and Sham M. Kakade and Dean Foster and Udaya Ghai},
  title = {{Mind the Gap: Examining the Self-Improvement Capabilities of Large Language Models}},
  year = {2024},
  note = {arXiv:2412.02674.}
}

@misc{suchman1987,
  author = {Lucy A Suchman},
  title = {{Plans and Situated Actions: The Problem of Human--Machine Communication}},
  year = {1987},
  note = {Cambridge University Press.}
}

@misc{suresh2021,
  author = {Harini Suresh and John V Guttag},
  title = {{A Framework for Understanding Sources of Harm throughout the Machine Learning Life Cycle}},
  year = {2021},
  doi = {10.1145/3465416.3483305},
  note = {In Proceedings of the 1st ACM Conference on Equity and Access in Algorithms, Mechanisms, and Optimization. ACM}
}

@misc{tian2023,
  author = {Katherine Tian and Eric Mitchell and Allan Zhou and Archit Sharma and Rafael Rafailov and Huaxiu Yao and others},
  title = {{Just Ask for Calibration: Strategies for Eliciting Calibrated Confidence Scores from Language Models Fine-Tuned with Human Feedback}},
  year = {2023},
  doi = {10.18653/v1/2023.emnlp-main.330},
  note = {In Proceedings of the 2023 Conference on Empirical Methods in Natural Language Processing. Association for Computational Linguistics, pages 5433-5442}
}

@misc{turpin2023,
  author = {Miles Turpin and Julian Michael and Ethan Perez and Samuel R. Bowman},
  title = {{Language Models Don’t Always Say What They Think: Unfaithful Explanations in Chain-of-Thought Prompting}},
  year = {2023},
  note = {In Advances in Neural Information Processing Systems 36.}
}

@misc{ustun2019,
  author = {Berk Ustun and Alexander Spangher and Yang Liu},
  title = {{Actionable Recourse in Linear Classification}},
  year = {2019},
  doi = {10.1145/3287560.3287566},
  note = {In Proceedings of the Conference on Fairness, Accountability, and Transparency. ACM, pages 10-19}
}

@misc{vasconcelos2023,
  author = {Helena Vasconcelos and Matthew Jörke and Madeleine Grunde-McLaughlin and Tobias Gerstenberg and Michael S. Bernstein and Ranjay Krishna},
  title = {{Explanations Can Reduce Overreliance on AI Systems During Decision-Making}},
  year = {2023},
  doi = {10.1145/3579605},
  note = {Proceedings of the ACM on Human-Computer Interaction 7(CSCW1): Article 129, 1--38}
}

@misc{veenman2006,
  author = {Marcel V. J. Veenman and Bernadette H. A. M. Van Hout-Wolters and Peter Afflerbach},
  title = {{Metacognition and Learning: Conceptual and Methodological Considerations}},
  year = {2006},
  doi = {10.1007/s11409-006-6893-0},
  note = {Metacognition and Learning 1(1): 3-14}
}

@misc{wachter2017,
  author = {Sandra Wachter and Brent Mittelstadt and Luciano Floridi},
  title = {{Why a Right to Explanation of Automated Decision-Making Does Not Exist in the General Data Protection Regulation}},
  year = {2017},
  doi = {10.1093/idpl/ipx005},
  note = {International Data Privacy Law 7(2): 76-99}
}

@misc{wachter2018,
  author = {Sandra Wachter and Brent Mittelstadt and Chris Russell},
  title = {{Counterfactual Explanations without Opening the Black Box: Automated Decisions and the GDPR}},
  year = {2018},
  note = {Harvard Journal of Law and Technology 31(2): 841-887.}
}

@misc{wang2023,
  author = {Xuezhi Wang and Jason Wei and Dale Schuurmans and Quoc V. Le and Ed H. Chi and Sharan Narang and others},
  title = {{Self-Consistency Improves Chain of Thought Reasoning in Language Models}},
  year = {2023},
  note = {In Proceedings of the Eleventh International Conference on Learning Representations.}
}

@misc{wang2024,
  author = {Yuqing Wang and Yun Zhao},
  title = {{Metacognitive Prompting Improves Understanding in Large Language Models}},
  year = {2024},
  doi = {10.18653/v1/2024.naacl-long.106},
  note = {In Proceedings of the 2024 Conference of the North American Chapter of the Association for Computational Linguistics: Human Language Technologies. Association for Computational Linguistics, pages 1914-1926}
}

@misc{wei2022,
  author = {Jason Wei and Xuezhi Wang and Dale Schuurmans and Maarten Bosma and Brian Ichter and Fei Xia and Ed H. Chi and Quoc V. Le and Denny Zhou},
  title = {{Chain-of-Thought Prompting Elicits Reasoning in Large Language Models}},
  year = {2022},
  doi = {10.52202/068431-1800},
  note = {In Advances in Neural Information Processing Systems 35, 24824--24837}
}

@misc{weston2023,
  author = {Jason Weston and Sainbayar Sukhbaatar},
  title = {{System 2 Attention: Is Something You Might Need Too}},
  year = {2023},
  note = {arXiv:2311.11829.}
}

@misc{xiong2024,
  author = {Miao Xiong and Zhiyuan Hu and Xinyang Lu and Yifei Li and Jie Fu and Junxian He and others},
  title = {{Can LLMs Express Their Uncertainty? An Empirical Evaluation of Confidence Elicitation in LLMs}},
  year = {2024},
  note = {In Proceedings of the Twelfth International Conference on Learning Representations.}
}

@misc{yang2024,
  author = {John Yang and Carlos E. Jimenez and Alexander Wettig and Kilian Lieret and Shunyu Yao and Karthik R. Narasimhan and Ofir Press},
  title = {{SWE-agent: Agent-Computer Interfaces Enable Automated Software Engineering. In Advances in Neural Information Processing Systems 37, 50528--50652}},
  year = {2024}
}

@misc{yao2023a,
  author = {Shunyu Yao and Dian Yu and Jeffrey Zhao and Izhak Shafran and Thomas L. Griffiths and Yuan Cao and Karthik R. Narasimhan},
  title = {{Tree of Thoughts: Deliberate Problem Solving with Large Language Models}},
  year = {2023},
  note = {In Advances in Neural Information Processing Systems 36, 11809--11822.}
}

@misc{yao2023b,
  author = {Shunyu Yao and Jeffrey Zhao and Dian Yu and Nan Du and Izhak Shafran and Karthik R. Narasimhan and Yuan Cao},
  title = {{ReAct: Synergizing Reasoning and Acting in Language Models}},
  year = {2023},
  note = {In The Eleventh International Conference on Learning Representations.}
}

@misc{yeung2012,
  author = {Nick Yeung and Christopher Summerfield},
  title = {{Metacognition in Human Decision-Making: Confidence and Error Monitoring}},
  year = {2012},
  doi = {10.1098/rstb},
  note = {Philosophical Transactions of the Royal Society B: Biological Sciences 367(1594): 1310-1321.  2011.0416}
}

@misc{yin2019,
  author = {Ming Yin and Jennifer Wortman Vaughan and Hanna Wallach},
  title = {{Understanding the Effect of Accuracy on Trust in Machine Learning Models}},
  year = {2019},
  doi = {10.1145/3290605.3300509},
  note = {In Proceedings of the 2019 CHI Conference on Human Factors in Computing Systems. ACM}
}

@misc{zhao2024,
  author = {Andrew Zhao and Daniel Huang and Quentin Xu and Matthieu Lin and Yong-Jin Liu and Gao Huang},
  title = {{ExpeL: LLM Agents Are Experiential Learners}},
  year = {2024},
  doi = {10.1609/aaai.v38i17.29936},
  note = {In Proceedings of the AAAI Conference on Artificial Intelligence. pages 19632-19642}
}

@misc{zhou2023,
  author = {Denny Zhou and Nathanael Schärli and Le Hou and Jason Wei and Nathan Scales and Xuezhi Wang and others},
  title = {{Least-to-Most Prompting Enables Complex Reasoning in Large Language Models}},
  year = {2023},
  note = {In Proceedings of the Eleventh International Conference on Learning Representations.}
}

@misc{zhou2024,
  author = {Yujia Zhou and Zheng Liu and Jiajie Jin and Jian-Yun Nie and Zhicheng Dou},
  title = {{Metacognitive Retrieval-Augmented Large Language Models}},
  year = {2024},
  doi = {10.1145/3589334.3645481},
  note = {In Proceedings of the ACM Web Conference 2024, 1453--1463. New York: Association for Computing Machinery}
}

\end{document}